\documentclass{article}
\usepackage{wrapfig} 
\usepackage[numbers]{natbib} 
\usepackage{authblk} 
\usepackage{orcidlink}
\usepackage{times}
\usepackage{amsmath}
\usepackage{amssymb}
\usepackage{amsfonts}
\usepackage{xspace}
\usepackage{textcomp}
\usepackage{stfloats}
\usepackage{float}
\usepackage{url}
\usepackage{verbatim}
\usepackage{graphicx}
\usepackage[inline]{enumitem}
\usepackage{url}
\usepackage{todonotes}
\presetkeys{todonotes}{inline,backgroundcolor=yellow,caption={}}{}
\graphicspath{ {./images/} }
\usepackage[export]{adjustbox}

\usepackage{stackengine,xcolor}
\usepackage{xspace}
\usepackage{algorithm}
\usepackage{algorithmicx}
\usepackage[noend]{algpseudocode}
\usepackage{listings}
\lstdefinestyle{PythonStyle}{
    language=Python,
    basicstyle=\small\ttfamily,
    numbers=left,
    numberstyle=\tiny,
    numbersep=5pt,
    tabsize=2,
    extendedchars=true,
    breaklines=true,
    keywordstyle=\color{blue},
    frame=b,
    stringstyle=\color{red},
    showspaces=false,
    showtabs=false,
    xleftmargin=17pt,
    framexleftmargin=17pt,
    framexrightmargin=5pt,
    framexbottommargin=4pt,
    showstringspaces=false
}

\lstdefinelanguage{NuXmv}{
    morekeywords={MODULE, VAR, ASSIGN, INIT, TRANS, SPEC, DEFINE, BOOLEAN, TRUE, FALSE, CTLSPEC, FAIRNESS, INVAR},
    sensitive=true,
    morecomment=[l]{--},
    morecomment=[s]{/*}{*/},
    morestring=[b]",
}

\lstdefinestyle{NuXmvStyle}{
    language=NuXmv,
    basicstyle=\small\ttfamily,
	numbers=left,
	numberstyle=\tiny,
	numbersep=5pt,
	tabsize=2,
	extendedchars=true,
	breaklines=true,
    keywordstyle=\color{blue},
    frame=b,
    stringstyle=\color{red},
    showspaces=false,
    showtabs=false,
    xleftmargin=17pt,
    framexleftmargin=17pt,
    framexrightmargin=5pt,
    framexbottommargin=4pt,
    showstringspaces=false
    commentstyle=\color{green!50!black},
    stepnumber=1,
    belowskip=10pt,
}

\lstdefinestyle{sqlstyle}{
    language=SQL,
    alsoletter={-},
    basicstyle=\ttfamily\scriptsize,
    keywordstyle=\color{blue},
    morekeywords={LIMIT-TO, TITLE-ABS-KEY, DOCTYPE, LANGUAGE, PUBYEAR, AFT},
    stringstyle=\color{green!50!black},
    commentstyle=\color{gray},
    showstringspaces=false,
    breaklines=true,
    frame=b,
    morestring=[b]",
    captionpos=b,
    numbers=left,
    numberstyle=\tiny,
    numbersep=5pt,
    tabsize=2,
    extendedchars=true,
    showspaces=false,
    showtabs=false,
    xleftmargin=17pt,
    framexleftmargin=17pt,
    framexrightmargin=5pt,
    framexbottommargin=4pt,
    showstringspaces=false,
    commentstyle=\color{green!50!black},
    stepnumber=1,
    belowskip=10pt,
    literate={LIMIT-TO}{{\textcolor{black}{LIMIT-TO}}}7
}

\usepackage{setspace}

\makeatletter
\algrenewcommand\ALG@beginalgorithmic{\footnotesize}
\makeatother
\usepackage{hyperref}
\UseRawInputEncoding
\usepackage{xspace}

\newcommand{\tab}{Table\xspace}
\def\sec{Section\xspace}

\definecolor{procColor}{HTML}{2ECC71}
\usepackage{multirow}
\usepackage{comment}

\usepackage[T1]{fontenc}
\usepackage{rotating}
\usepackage{tabularx}
\usepackage{hyperref}
\usepackage{caption}
\usepackage{threeparttable}
\usepackage{booktabs}
\begin{document}

\title{SoK: Concurrency in Blockchain - A Systematic Literature Review and the Unveiling of a Misconception}

\date{}

\author[1]{Atefeh Zareh Chahoki\orcidlink{0009-0003-2004-7762}}
\author[2]{Maurice Herlihy \orcidlink{0000-0002-3059-8926}}
\author[1]{Marco Roveri \orcidlink{0000-0001-9483-3940}}

\affil[1]{University of Trento, Trento, Italy\\
          \texttt{atefeh.zareh@unitn.it}, \texttt{marco.roveri@unitn.it}}
\affil[2]{Brown University, Providence RI, USA\\
          \texttt{mph@cs.brown.edu}}

\maketitle

\begin{abstract}
  Smart contracts, the cornerstone of blockchain technology, offer the prospect of secure and automated distributed execution. Given their involvement in managing vast volumes of transactions across various roles-including clients, miners, and validators- exploring concurrency is pivotal from multiple perspectives. This includes executing or validating transactions concurrently within each block, concurrent block processing across different shards, and concurrent competition among miners to select and persist transactions within blocks. Concurrency and parallelism represent a double-edged sword: Although increasing concurrency at any level benefits higher throughput, it also introduces potential risks such as race conditions, non-deterministic outcomes, and vulnerabilities such as deadlock and livelock. 

This paper presents the first survey of concurrency in smart contracts, offering a comprehensive systematic review of the literature, and organizing it into distinct dimensions. 
Firstly, it establishes a comprehensive taxonomy of existing levels of concurrency within blockchain systems and discusses currently proposed solutions for incorporation into future blockchain iterations. 
Secondly, given the imperative for correctness and security in massively distributed processing infrastructures, this paper examines concurrent operations' vulnerabilities, potential attacks, and countermeasures. 
Critically, we analyze the assumptions used by each category and highlight a flaw in the concurrency assumption in a major category. This misconception has led to misinterpretations in important research bodies. Thus, this work aims to correct long-standing misconceptions within the field and steer future research toward more genuine assumptions. Finally, we clarify the directions for future research by identifying gaps within each category, thus contributing to the advancement of the blockchain community.
\end{abstract}

\maketitle

\section{Introduction}
Smart contracts are programs that can be stored and executed on a blockchain and that automate the execution of agreements according to predefined logic between untrusted parties~\cite{next_buterin_2014}. Due to their tamper-proof nature and immutability of state transitions, they are ideal for enabling and facilitating trustless transactions in various domains. Deterministic execution is essential for smart contracts to ensure the predictable outcome of a transaction based solely on its inputs and the current state of the blockchain. This predictability is crucial for reliable decision-making and for avoiding unintended consequences.
The concept known as the "Blockchain impossibility triangle"~\cite{Long_Brief_2021} or "Blockchain Trilemma"~\cite{Conti_Blockchain_2019} suggests that a blockchain system cannot simultaneously achieve optimal scalability, decentralization, and security. Although this trilemma is not a formal proof within blockchain architecture, it effectively encapsulates the primary challenges blockchain systems face as observed in current implementations and industry practices. No blockchain system excels in all three areas. Traditionally, most blockchain systems have prioritized decentralization and security over scalability \cite{Zhang_Towards_2018}, which fundamentally limits their performance. Similarly to how different distributed systems must balance consistency and availability due to the CAP theorem, blockchain systems must also navigate trade-offs between scalability, decentralization, and security depending on the scenario. Scalability should not always be treated as the least important factor and sacrificed as a result.

One approach to addressing the scalability challenge is concurrency. This concept can be applied at various levels of the blockchain infrastructure, such as sharding, which enables transactions to be processed concurrently across different shards, or executing smart contracts within each shard and for each block by validators concurrently in a multi-core setting. Other strategies explored in this survey also leverage concurrency to enhance scalability. 

However, concurrency introduces challenges related to synchronization, consistency, race conditions, deadlocks, and livelocks. These challenges can result in security vulnerabilities and unexpected behavior in smart contracts. Therefore, ensuring that these issues are addressed is critical for any concurrency solution, as discussed in this survey.

This survey aims to provide a comprehensive and systematic review of all types of concurrency issues in smart contracts. It assesses the validity of the assumptions and breaks them down into a taxonomy and a more detailed organization. In each area, the future open questions are elaborated on and presented. The study also aims to examine weaknesses, attacks, and countermeasures in the field, critically analyze assumptions, and correct long-held misconceptions to align future research with real assumptions.

This paper presents several key contributions: \begin{enumerate*}[label=(\arabic*)] \item To the best of the authors' knowledge, it provides the first comprehensive survey of concurrency in smart contracts, offering a systematic review of a wide range of current literature to serve as an organized and insightful resource for researchers; \item a detailed taxonomy is introduced, classifying various concurrency aspects into categories and organizing them based on their approaches and goals, whether implemented on existing blockchains or proposed for future developments; \item open questions and emerging trends in all defined categories are highlighted, offering valuable guidance for future research; and \item the flawed assumption of concurrent execution of transactions in Category Six of the presented taxonomy is critically examined and discussed, supported by sufficient references.\end{enumerate*}  All materials and code required to conduct this survey are publicly available on GitHub~\footnote{https://github.com/Atefeh-Zareh/concurrency-in-smart-contracts}.

The remainder of this paper is organized as follows: \sec{}~\ref{sec:Background} provides the main background concepts used throughout the paper. The methodology of the systematic review is presented in \sec{}~\ref{sec:Methodology}. \sec{}~\ref{sec:Taxonomy} organizes the concurrency perspectives into three categories. In \sec{}~\ref{sec:FlawedAssumption} a critical analysis reveals a misunderstanding in the third category. 

\section{Problem Statement And Motivation}\label{Problem_Statement_And_Motivation}
Existing blockchain systems require that every node in the network maintain a complete replication of the transaction history and ledger states to ensure the total order of transactions, execution integrity, and data provenance. These data structures grow significantly over time, becoming prohibitively large. For instance, as of July 2024, the Bitcoin ledger is approximately 550GB \cite{noauthor_blockchaincom_2024}, while the Ethereum ledger exceeds 18.5TB \cite{etherscanio_Archive_ethereum_2024}. To address this, most blockchain nodes maintain a smaller and compact index known as \textit{validation states}, which are sufficient to validate transactions. However, even these validation states can be substantial in size, such as Ethereum'cas validation states, which are around 1,120GB\cite{etherscanio_Archive_ethereum_2024}. Additionally, blockchain nodes must locally replay all transactions based on these replicated states, incurring significant storage and computation costs that limit system scalability. This cumbersome stateful data also undermines system security and robustness by centralizing the network, as fewer nodes can handle such large data volumes.

To mitigate these issues, one approach is sharding, which partitions the blockchain into multiple parallel chains, each managed by a subset of nodes. This reduces storage and computation duplications among different shards. However, sharding only alleviates the problem by a constant factor, as nodes within each shard still duplicate storage and computation. Moreover, sharding introduces new challenges, such as cross-shard transaction processing and vulnerability to attacks by slowly adaptive Byzantine adversaries. Shardening is studied in \sec{}~\ref{ssec:category-sharding}.

Another approach is the use of light nodes that store only block headers rather than full states. Although light nodes can follow consensus protocols, they cannot independently verify transaction validity, failing to address centralization due to state maintenance burdens.

The stateless blockchain concept has recently emerged as an off-chain scaling approach. This method moves ledger states and transaction executions off-chain to a subset of nodes, thereby reducing on-chain load. However, existing stateless blockchain systems are primarily designed for cryptocurrencies and face limitations when applied to general-purpose use cases involving smart contracts. Developing a general-purpose stateless blockchain that supports smart contracts presents several challenges. Transactions involving smart contracts can contain arbitrary logic, requiring novel proof techniques to ensure the integrity of off-chain executions. Smart contract transactions also introduce read-and-write sets of varying sizes, necessitating additional design for on-chain commitment updates. Furthermore, existing methods for cryptocurrency exchange offer limited support for parallel transaction executions. To enhance system throughput, new transaction processing methods are needed to validate and commit concurrent transactions from an asynchronous network, despite the stateless design. This category of off-chain solutions is studied in \sec{}~\ref{ssec:category-offchain}.

\section{Background}\label{sec:Background}
This section briefly reviews the concepts related to concurrency and blockchain used in this paper.
\subsection{Concurrency concepts}\label{ssec:Background_concurrency}

Concurrency refers to the ability of a computer program to execute multiple instructions or processes simultaneously. In distributed systems, concurrency can introduce challenges related to:
\begin{itemize} 
    \item \textbf{Consistency}: Maintaining data integrity across different replicas or copies in a distributed system.
    
    \item \textbf{Synchronization}: Ensuring that multiple processes access shared resources in a coordinated manner to avoid data inconsistencies.
    
    \item \textbf{Race Conditions}: Scenarios where the outcome of a program depends on the unpredictable timing of concurrent processes.
   
    \item \textbf{Nondeterminism}: Parallel systems commonly demonstrate this trait due to competition for communication resources. A system shows nondeterminism when two identical copies of it can act differently even with the same inputs upon which wins the race that results in an untestable trait of these systems.~\cite{Theory_Roscoe_1998}
    
    \item \textbf{Deadlock}: is one of the most common issues in a parallel system, when no component can progress due to mutual waiting for communication, a concurrent system is deadlocked. A classic example is the 'five dining philosophers', where each philosopher requires both neighboring forks to eat, leading to a deadlock if all philosophers simultaneously pick up their left-hand fork and they deadlock and starve to death. One of the main reasons for deadlock is competition to acquire the resources~\cite{Theory_Roscoe_1998}. 
    \item \textbf{Livelock}: The cause of livelock, is called \textit{divergence} which means a program performs infinite unbroken sequences of internal actions and never interacts with its environment, such as infinite loops in programs. Although operationally and theoretically deadlock and livelock are different, they appear similar from the users' perspective because no response is received. However, livelock is worse since users may observe internal activities and mistakenly expect an eventual output~\cite{Theory_Roscoe_1998}.
\end{itemize}

\subsection{Smart contracts}\label{ssec:Background_Smart_Contract}
The \textit{blockchain} is a foundational design pattern to facilitate pure peer-to-peer distributed computation. Initially implemented by Bitcoin in 2009 as a cryptocurrency~\cite{nakamoto_bitcoin_2008}, it has since evolved into a robust infrastructure capable of executing a wide range of generalized functionalities beyond cryptocurrency. Ethereum~\cite{buterin_ethereum_white_2013}, in particular, pioneered this expansion in \citeyear{buterin_ethereum_white_2013} by introducing the \textit{Ethereum Virtual Machine (EVM)}, a Turing-complete state machine that handles both deployment and execution of codified arbitrary business logic scripts named \textit{smart contracts}. All blockchain networks maintain securely shared data named \textit{ledger} through a \textit{consensus protocol} facilitated by volunteer operators named \textit{miners} and \textit{validators}. The transactions announced by users are initially stored in the \textit{mempool} (short for "memory pool"), a temporary storage area where pending transactions wait to be included in the next block. Miners select transactions from the mempool that offer higher fees first, as this maximizes their rewards and persists them into the next block if the transaction is valid. This validity check includes the correctness of transaction format, having a sufficient gas fee, nonce verification, signature verification, balance check, double-spending prevention, compliance with consensus rules, and regarding smart contracts ensuring that the transaction can execute without error. Then miners extend the blockchain by one block in each \textit{block interval}, which varies across different networks such as 10 minutes in Bitcoin or 12 seconds in Ethereum. Smart contracts introduce the concept of \textit{gas}, irrespective of the consensus protocol they utilize. Gas represents the computational effort and cost required to execute operations within smart contracts to compensate for the resources used to execute smart contracts. It prevents infinite loops, restricts resource consumption, and maintains the network's efficiency and security. Smart contracts are written in various programming languages tailored to specific blockchain platforms. Solidity is the predominant language for Ethereum; Vyper aims to provide simplicity and security; Michelson is used on Tezos; and Move was developed for Diem (formerly known as Libra).

\subsection{Proof of Work (PoW) \& Bitcoin}\label{ssec:Background_PoW}
Proof of Work (PoW) is a consensus protocol that is mainly introduced in Bitcoin. Participants, known as \textit{miners}, gather newly broadcasted and not yet persisted transactions and add them in a \textit{block} data structure if they are \textit{valid}, then compete to solve a cryptographic puzzle, named \textit{mining}, to add the block to the ledger. One part of the block is its \textit{previous block hash}, which makes this data structure \textit{tamper-proof}. The first miner to solve the puzzle earns the right to add the next block of transactions to the blockchain and win rewards. Rewards are in the form of the newly \textit{minted} cryptocurrencies in that block and the \textit{fee} and gas of the included transactions as the incentive. PoW is the original consensus mechanism used by Bitcoin~\cite{nakamoto_bitcoin_2008} and several other cryptocurrencies, but because it is energy-intensive, it is being replaced by others such as Ethereum.

\subsection{Proof of Stake (PoS) \& Ethereum Mechanism}\label{ssec:Background_PoS}
This section provides an overview of Proof of Stake (PoS), one of the most widely adopted consensus algorithms in blockchain infrastructures, including Ethereum. This section also briefly explores Ethereum's execution mechanism, highlighting its key components and processes.

In PoS, a node is selected as the proposer using methods like computational contests, predetermined sequences, or random algorithms, as seen in Ethereum. The proposer compiles a block by selecting a set of transactions and shares it with other nodes, known as attestors. Each attestor independently validates the transactions in the block, rejecting it if any discrepancies or invalid transactions are found.

Ethereum transitioned to PoS with its Ethereum 2.0 upgrade in 2022, aiming to enhance security, reduce energy consumption, and support new scaling solutions compared to the prior Proof of Work (PoW) architecture. In PoS, participants, referred to as \textit{validators} or \textit{nodes}, must deposit 32 ETH into a deposit contract, thereby \textit{staking} it as collateral~\cite{noauthor_PoS_2024}. Validators are responsible for ensuring the validity of new blocks and occasionally for proposing and propagating them. Misbehavior, such as proposing multiple conflicting blocks or sending invalid attestations, results in penalties, including partial or total loss of staked ETH. To join as a validator, a user must pass through an activation queue, which limits the rate of new validators joining the network. This mechanism ensures system stability, efficient consensus, and protection against resource strain and Sybil attacks, where a single entity creates multiple identities to gain disproportionate influence over the network.

In Ethereum's PoS system, block timing follows a fixed schedule rather than being influenced by mining difficulty as in PoW. Time is divided into \textit{slots} of 12 seconds and \textit{epochs} comprising 32 slots, corresponding to 6.4 minutes. During each slot, a validator is pseudo-randomly selected using RANDAO to propose a new block and broadcast it to the network. Simultaneously, a randomly chosen committee of validators evaluates the proposed block's validity. These committees distribute the workload, ensuring that all active validators participate within each epoch without needing to act in every slot~\cite{noauthor_PoS_2024}.

\subsubsection{Transaction Process}

The process of executing a transaction in Ethereum PoS involves several steps. A user creates and signs a transaction with their private key via a wallet or library, essentially making a request to a node through the \texttt{Ethereum JSON-RPC API}. The transaction is submitted to the \textit{execution client} part of that node, which verifies its validity, ensuring that the sender has sufficient ETH and has signed the transaction correctly. Valid transactions are added to the \textit{local mempool} (a list of pending transactions) and broadcast over the execution layer gossip network. Advanced users may bypass public broadcasting and forward their transactions to specialized block builders, such as Flashbots Auction, to maximize profits through \textit{Maximal Extractable Value (MEV)}~\cite{noauthor_PoS_2024}.

The selected validator for the current slot is responsible for updating the global state. Nodes run three key software components: the \textit{execution client}, the \textit{consensus client}, and the \textit{validator client}. The execution client collects transactions from the local mempool, executes them locally to generate state changes, and compiles them into an \textit{execution payload}. This payload is passed to the consensus client, which incorporates it into a \textit{beacon block}. The beacon block also contains metadata such as rewards, penalties, slashings, and attestations, enabling consensus on the blockchain's state and head of the chain~\cite{noauthor_PoS_2024}.

Other nodes receive the new beacon block via the consensus layer gossip network and pass it to their execution clients for local validation of the proposed state changes. The validator client attests to the block's validity and confirms that it builds upon the chain with the greatest attestation weight, as determined by the fork choice rules. Once validated, the block is added to the local database of each node.

\subsubsection{Transaction Finality}
A transaction achieves \textit{finality} when it is part of a block that cannot be reverted without burning a significant amount of staked ETH. Finality is managed through \textit{checkpoint} blocks, which occur at the start of each epoch. Validators vote on checkpoint pairs, and if a pair receives votes from at least two-thirds of the total staked ETH, the target checkpoint becomes \textit{justified}, and the source checkpoint is upgraded to \textit{finalized}. Reverting a finalized block would require an attacker to control and sacrifice at least one-third of the total ETH staked across all participants in the network. To prevent an attacker from stalling finality, Ethereum employs an \textit{inactivity leak}, which penalizes validators voting against the majority when finality fails for more than four epochs. This ensures the majority regains a two-thirds stake to finalize the chain.

\subsubsection{Ethereum's Layered Architecture}
Ethereum's architecture is organized into several layers, each responsible for specific functions that contribute to the overall operation and scalability of the network. The \textbf{execution layer} handles transaction processing, smart contract execution, and state changes. This layer is responsible for bundling transactions into \textit{execution blocks} and updating the global state. Above the execution layer is the \textbf{consensus layer}, also known as the \textit{Beacon Chain}, which is responsible for maintaining network consensus, coordinating validators, and ensuring security. The \textit{beacon block} is the key data structure in this layer, as it organizes validator attestations and records the consensus state of the network. Ethereum also integrates a \textbf{data availability layer}, which ensures the reliable retrieval of off-chain data, particularly for layer 2 solutions. Lastly, the network utilizes \textbf{layer 2 scaling solutions}, such as \textbf{Optimistic Rollups} and \textbf{zkRollups}, which offload transaction processing to achieve greater throughput while maintaining security through periodic anchors to the mainnet. This modular, layered architecture enhances Ethereum's scalability and security while enabling the development of decentralized applications and solutions.

\subsubsection{Ethereum Networks and Their Purposes}
Ethereum operates across several networks, each tailored for specific purposes within its ecosystem. The \textbf{mainnet} is the primary public network where real transactions, decentralized applications (dApps), and smart contracts are executed, using ETH with actual monetary value. For development and testing, Ethereum offers \textbf{testnets} such as \textbf{Goerli} and \textbf{Sepolia}, which simulate mainnet conditions without risking real assets, with Goerli widely adopted for advanced testing and Sepolia preferred for lightweight development. \textbf{Private networks} are custom deployments used by organizations or developers for controlled experimentation and application testing. To address scalability, Ethereum integrates \textbf{layer 2 networks} like \textbf{Optimism} and \textbf{Arbitrum} (Optimistic Rollups) and \textbf{zkSync} and \textbf{StarkNet} (zkRollups), which reduce transaction costs and increase throughput by offloading computation from the mainnet while retaining security guarantees. Additionally, emerging \textbf{data availability solutions} like \textbf{EigenLayer} ensure that off-chain data for rollups can be reliably retrieved, further enhancing Ethereum's scalability and modularity.

\subsubsection{Terminology and Roles in Ethereum PoS}\label{ssec:Background_Terminology_Roles_Ethereum}
In Ethereum's PoS architecture, the terms \textit{validator} and \textit{node} are used interchangeably. Each node performs multiple roles: proposing blocks when pseudo-randomly selected for a slot, attesting to other validators' blocks within an epoch as part of a committee, and validating transactions in each slot. These roles are not distinct among nodes but represent different responsibilities within each node.

However, various terminologies in the literature might appear to suggest different roles. For instance, some references distinguish \textit{miners} and \textit{validators}, where miners propose new blocks and validators verify them~\cite{Adding_Dickerson_2020}. Others, such as \cite{Efficient_Xia_2023}, use terms like \textit{proposer} and \textit{attestor} within a proposal-attestation paradigm. In this context, proposers select transactions and broadcast blocks, while attestors validate the blocks. In Ethereum, these distinctions are purely functional, as all nodes perform these tasks at different times.

In this survey, we adhere to the Ethereum convention, referring to participants as \textit{validators}. For consistency and clarity, references to other terminology will include appropriate explanations, ensuring that readers are not misled by differing conventions.

\subsection{Hyperledger Fabric}\label{ssec:Background_Hyperledger_Fabric}

\subsubsection{Nodes in Hyperledger Fabric}

In Hyperledger Fabric, the nodes serve as communication entities within the blockchain network. Unlike many blockchains where nodes operate in a peer-to-peer manner, nodes in Hyperledger Fabric have distinct roles. There are three types of nodes:

\begin{itemize}
    \item \textbf{Client}: The Client is operated by the end user and is responsible for submitting transaction proposals to the Endorser Peer and broadcasting the proposals and responses to the Orderer.
    
    \item \textbf{Peer}: Peers are primarily tasked with executing chaincode to read from and write to the ledger. All Peers act as committing peers (Committers), maintaining the ledger's state. Additionally, peers can serve as endorsing peers (endorsers) when an application makes a transaction endorsement request. This endorsement role is dynamic; otherwise, Peers function as regular committing peers.
    
    \item \textbf{Orderer}: The Orderer nodes collectively form the ordering service. In Hyperledger Fabric, which is a distributed system, each node maintains a copy of the ledger. The ordering service ensures the consistency of the ledger by ordering transactions into blocks.
\end{itemize}

\subsection{Other Consensus Protocols}\label{sssec:Consensus_Protocols}

There are two categories of consensus protocols:
Proof-based and Leader-based~\cite{Zhang_Towards_2018}.

\textit{Byzantine fault tolerance (BFT)} is a critical property in distributed system design. It refers to the system's ability to maintain its correct operation even when some components fail or behave maliciously. In blockchain networks, Consensus protocols are mechanisms used by distributed systems to achieve agreement on a single data value among multiple nodes or participants, even in the presence of faulty or malicious actors. This property is essential to maintain the integrity and security of the distributed ledger. Various consensus algorithms have been developed to achieve Byzantine fault tolerance in blockchain systems, each with its own trade-offs in terms of efficiency, scalability, and security. The following reviews some of the main types of consensus protocols, excluding PoW and PoS, which are discussed later~\cite{Lashkari_Comprehensive_2021}:

\begin{itemize}
 \item \textbf{Delegated Proof of Stake (DPoS):} DPoS is a variation of PoS where token holders vote for a select few delegates who are responsible for validating transactions and adding them to the blockchain. EOS and TRON are examples of blockchain platforms that use DPoS.
 
 \item \textbf{Proof of Authority (PoA):} In PoA, consensus is achieved by a set of approved validators, typically known entities or organizations. Validators are responsible for creating new blocks and maintaining the blockchain. PoA is used in networks where a high degree of trust among participants is assumed, such as private or consortium blockchains.

\item \textbf{Practical Byzantine Fault Tolerance (PBFT):} PBFT is a consensus algorithm designed to work in asynchronous networks with a certain number of faulty or malicious nodes. It ensures that all honest nodes reach a consensus on the order of transactions. PBFT is used in permissioned blockchains and distributed systems like Hyperledger Fabric.
 
 \item \textbf{Raft:} Raft is a consensus algorithm designed for fault-tolerant distributed systems. It elects a leader node responsible for managing the replication of the log among other nodes. Raft is often used in distributed databases and systems where simplicity and ease of understanding are prioritized.
 
 \item \textbf{Directed Acyclic Graph (DAG):} DAG-based consensus protocols, such as Tangle (used by IOTA) and Hashgraph, do not rely on blocks or miners. Instead, transactions are directly linked to each other in a graph-like structure, and consensus is achieved through a voting process based on transaction approval.

\item \textbf{Proof of Vote (PoV):} Proof of Vote (POV)~\cite{Proof_Li_2017} represents a consensus algorithm designed specifically for consortium blockchains where PoW falls short. In this model, consensus is orchestrated by distributed nodes overseen by consortium partners, reaching decentralized decisions through voting-based arbitration. The primary concept involves assigning distinct security identities to network participants, enabling block submission and verification via voting among these entities, eliminating the need for third-party intermediaries or reliance on unregulated public awareness. In contrast to the fully decentralized PoW consensus, POV offers controllable security, convergence, and reliability, achieves transaction finality with a single block confirmation, and ensures swift transaction verification with minimal delay.
\end{itemize}

\subsection{Performance in smart contracts}

Table~\ref{tab:TransactionSpeed} shows different cryptocurrencies in the first column, the transaction speed as TPS in the second column, which stands for "transactions per second" and the last column shows the "average transaction confirmation time" of these cryptocurrencies~\cite{Improving_Hazari_2020}. The transaction verification process for cryptocurrencies is very slow and does not match the performance of traditional payment systems such as VISA, which handles an average of 10,547 TPS and can peak at 56,000 TPS and Paypal with a TPS of 1,700. TPS for Bitcoin is almost 7 and other TPSs for cryptocurrencies are listed in the second column on this paper. The calculation of the TPS for the cryptocurrencies in the last year is available in this paper git path. 

Numerous approaches have been proposed to increase transaction speed and scalability by enhancing concurrency efficiency in blockchain technology. In Section \ref{sec:Taxonomy}, we will review these approaches, focusing on how concurrency is utilized at different levels of the blockchain to achieve improved performance.
 
\begin{center}
 \captionsetup{type=table}
\captionof{table}{transaction speed of different cryptocurrencies}
\label{tab:TransactionSpeed}
\begin{tabular}{lll}
\hline
\textbf{Cryptocurrency} 
&
\textbf{TPS} 
&
\textbf{Time} 
\\ \hline
Bitcoin     	& $3-7$     & $60 \mathrm{~min}$	\\
Ethereum    	& $15-25$   & $6 \mathrm{~min}$		\\
Ripple      	& 1500      & $4 \mathrm{~sec}$		\\
Bitcoin Cash	& 61        & $60 \mathrm{~min}$	\\
Stellar     	& 1000     & $2-5 \mathrm{~sec}$	\\
Litecoin    	& 56        & $30 \mathrm{~min}$	\\
Monero      	& 4         & $30 \mathrm{~min}$	\\
IOTA         	& 1500      & $2 \mathrm{~min}$	\\
Dash        	& 48        & $2-10 \mathrm{~min}$	\\ \hline
\end{tabular}
\end{center}

\section{Methodology} \label{sec:Methodology}
This section outlines the questions to be addressed, along with the source dataset and the query used to extract the relevant literature for this systematic review. Subsequently, it presents the exclusion criteria. 

\subsection{Research questions} \label{ssec:Research_Questions}
This study aimed to address several crucial research inquiries:

\textbf{(RQ1)}: What are the distinct concurrency aspects associated with smart contracts and their blockchain infrastructure?\\
- This question focuses on gaining insights into all types of concurrency aspects within the blockchain ecosystem and organizing them into categories.\\

\textbf{(RQ2)}: What are the existing security threats and countermeasures available in the domain of blockchain concurrency, and how do they address the identified concerns?\\
- This investigation aims to identify the techniques currently utilized to address concurrency challenges in smart contracts in a detailed representation. It categorizes the literature based on whether the aspect already exists in the blockchain infrastructures or is proposed for further iterations. In addition, it examines whether the study's focus is on introducing vulnerabilities, attacks, or security solutions, and analyzes the approaches taken to address these issues.\\

\textbf{(RQ3)}: What are the potential research gaps of current solutions that require attention?\\
- The purpose of this inquiry is to organize the open questions and trends in all these categories. These identified threats can serve as guidelines for future research in this field, directing attention to areas that require further investigation due to the absence of proposed solutions or the limitations of existing approaches.\\

\subsection{Source databases and search query} \label{ssec:Source_Databases}
This study used Scopus\footnote{\texttt{\url{https://www.scopus.com/}}} as the primary source to identify relevant publications in this field. Scopus provides a vast repository of scientific literature with over 27,950 active titles, 90.6+ million records, and from more than 7,000 publishers. Scopus is continuously expanding its coverage of conference material primarily for the subject domains of publishers and societies of engineering and computer sciences~\cite{noauthor_scopus_2024}. In addition, Scopus offers a powerful search engine specifically designed for academic exploration. Although other databases such as Google Scholar, Web of Science, and IEEE Xplore exist, the extensive coverage and robust search capabilities of Scopus make it a sufficient choice for this particular research.

The search query utilized for all repositories can be found in Listing \ref{lst:search_query}:

\begin{figure}[h!] 
\begin{lstlisting}[style=sqlstyle, caption={Search query.}, , firstnumber=1, label={lst:search_query}, escapechar=\%]
TITLE-ABS-KEY ( %\label{line:searchType}%
  ( "smart contract" OR "Ethereum" OR "solidity" OR "Hyperledger" ) %\label{line:blockchain}%
  AND ( "concurren*" OR "race" OR "parallel" OR 
        "multithreaded" OR "synchroni*" OR "transaction ordering dependancy" OR "front running attack" OR shardening) %\label{line:concurrency}%
  AND ( LANGUAGE( "English" ) ) %\label{line:language}%
  AND NOT( DOCTYPE ( "cr" ) OR  DOCTYPE ( "ab" ) OR DOCTYPE ( "bk" ) OR DOCTYPE ( "ch" ) OR DOCTYPE ( "cr" ) OR DOCTYPE ( "re" ) OR DOCTYPE ( "sh" ) )%\label{line:doctype}%
\end{lstlisting}
\end{figure}

In line \ref{line:searchType}, TITLE-ABS-KEY indicates that the search query is applied to all the title, abstract, and keyword sections of the repositories. The primary objective of the search is to identify research works that specifically address concurrency solutions \ref{line:concurrency}) in blockchain infrastructures (line \ref{line:blockchain}). The query utilized the wildcard * in 'concurren*' and 'synchroni*' to encompass all variations within the word families of concurrency and synchronization. The query targeted all papers in the English language (line \ref{line:language}) and excluded document types of conference review (cr), abstract report (ar), book (bk), book chapter (ch), conference review (cr), review (re) and short survey (sh), respectively, encoded on line \ref{line:doctype}. The described query execution in May 2024 returned 237 unique papers before the pruning process.

\subsection{Inclusion and exclusion criteria}\label{ssec:Exclusion_Criteria}
\tab{}~\ref{tab:InclusionExclusion} lists the inclusion and exclusion criteria that are applied in this current study.

\begin{center}
\captionsetup{type=table}
\captionof{table}{Inclusion and Exclusion Criteria}
\begin{tabular}{>{\raggedright\arraybackslash}p{0.45\columnwidth}>{\raggedright\arraybackslash}p{0.45\columnwidth}}
\hline
\textbf{Inclusion Criteria} & \textbf{Exclusion Criteria} \\ \label{tab:InclusionExclusion}
\\ \hline
\begin{itemize}[left=0pt,itemsep=0pt,parsep=0pt,topsep=0pt,partopsep=0pt]
  \item Research studying any concurrency aspects in smart contracts.
  \item Studies that address at least one of the designated research inquiries.
  \item Relevant studies identified through snowballing techniques from previously chosen literature.
\end{itemize}
& 
\begin{itemize}[left=0pt,itemsep=0pt,parsep=0pt,topsep=0pt,partopsep=0pt]
  \item Theses, books, survey-based studies, or patents.
  \item Non-peer-reviewed research.
  \item Duplicate studies.
  \item Studies lacking relevance to the designated research questions.
\end{itemize} \\
\hline
\end{tabular}
\end{center}

\subsection{Screening Results}\label{ssec:Screening_Results}
\begin{figure}[ht]
\centering
\includegraphics[width=0.5\textwidth]{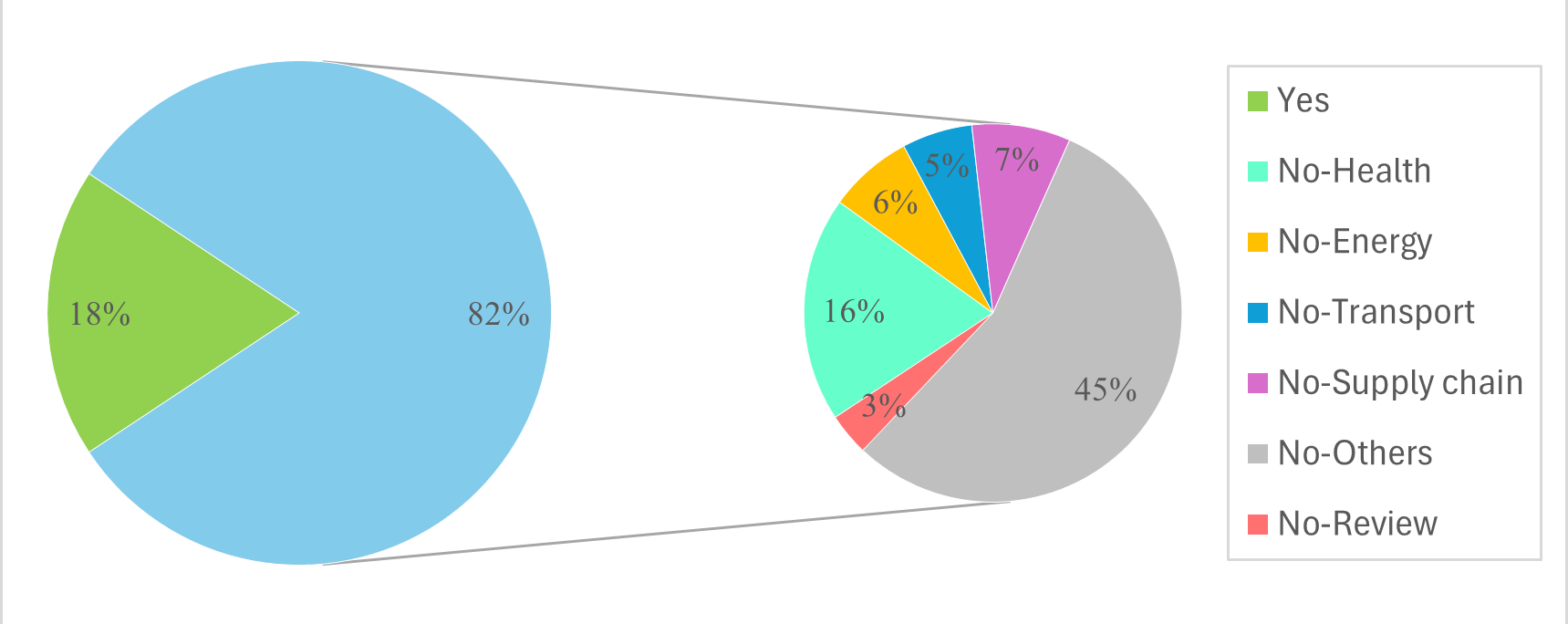}
\caption{Screening results}
\label{fig:screening-results}
\end{figure}

Figure \ref{fig:screening-results} illustrates the percentage of relevant articles discovered during the screening process. Additionally, a sub-pie chart categorizes the non-relevant papers into different domains, highlighting the areas that were identified and filtered out from the main set of papers under consideration.

\begin{figure}[ht]
\centering
\includegraphics[width=0.5\textwidth]{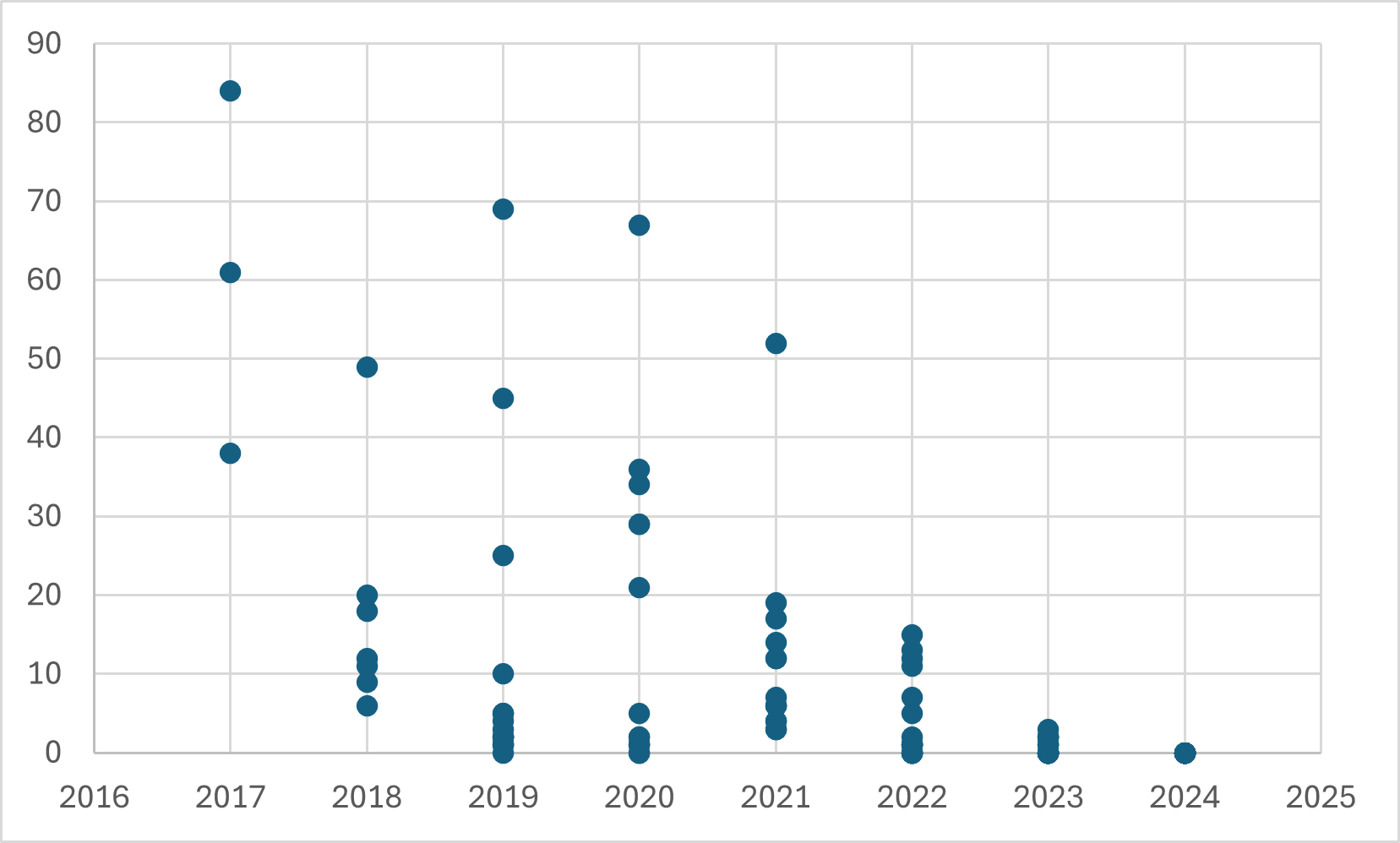}
\caption{Paper citations per year}
\label{fig:paper-citations}
\end{figure}

Figure \ref{fig:paper-citations} depicts the total number of citations for the relevant articles, categorized by their year of publication.

\begin{figure}[ht]
\centering
\includegraphics[width=0.5\textwidth]{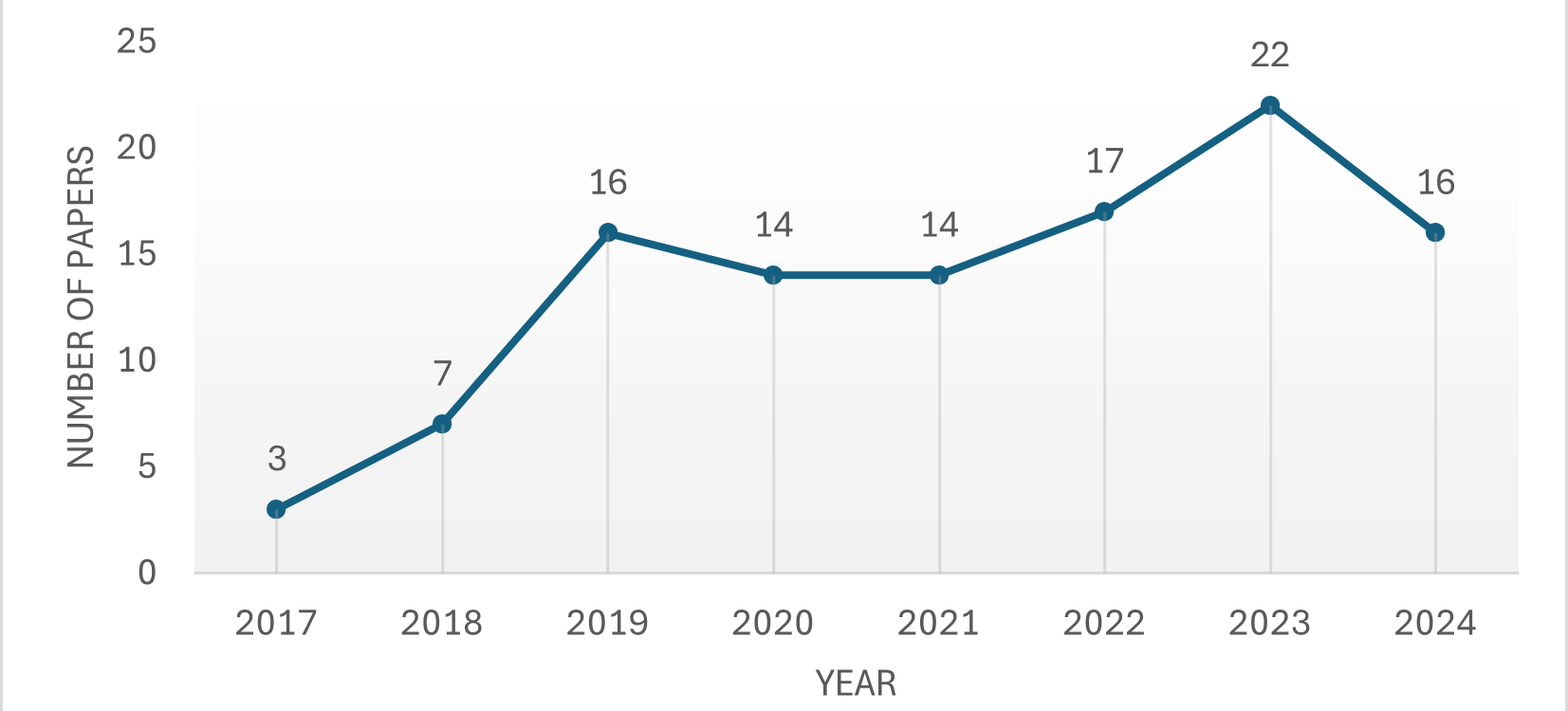}
\caption{Paper number per year}
\label{fig:paper-numbers-per-year}
\end{figure}

Figure \ref{fig:paper-numbers-per-year} shows the number of relevant articles published each year.

\section{Concurrency in smart contracts: a taxonomy}\label{sec:Taxonomy}
Three main categories of concurrency approaches for smart contracts have emerged that are represented in this section.

\subsection{Category 1: Concurrent Consensus}\label{ssec:category-Consensus}
This category of solutions focuses on consensus protocols and believes that to scale blockchain systems, the fundamental structure of consensus needs modification. They propose novel consensus protocols to enhance concurrency in blockchain infrastructure aiming to improve performance.

One notable research in this category is done by \citeauthor{parallel_Hazari_2019}~\cite{parallel_Hazari_2019}, who introduced \textit{parallel PoW} based on parallel mining rather than solo mining. This approach eliminates simultaneous efforts of miners to solve a specific block by introducing a new role as a \textit{manager}, along with processes for manager selection, work distribution, and a reward system. The introduction of a manager role does not compromise the decentralized nature of blockchain. This solution was implemented in a test environment mimicking Bitcoin's PoW features and tested across various scenarios with different difficulty levels and numbers of validators. The results indicate a 34\% improvement in PoW scalability. This protocol aims to increase transaction processing speed, potentially reducing energy consumption. This research is extended by these authors in \cite{Improving_Hazari_2020} which evaluated a cloud-based environment. The future plans for this research include evaluation against the 51\% attack, evaluation on the Bitcoin testnet, refinement of the reward system, and evaluation of real-time network energy consumption, although it is expected to end reduced energy consumption due to speeding up transaction processing.

\cite{Parallel_Liu_2022} introduces an approach to execute smart contracts in parallel and asynchronously. Unlike traditional methods, this approach minimizes coordination efforts and eliminates risks such as livelocks, even with small groups. This paradigm is implemented in two ways: first, enhancing Ethereum to support parallel and asynchronous execution seamlessly, without the need for hardforks. Second, introducing SaberLedger, a new public and permissionless blockchain, demonstrating its capabilities through a prototype implementation.

In \cite{BDLedger_Huang_2021}, the authors introduce BDLedger (Big Data Ledger), a blockchain-like distributed ledger designed to achieve a nearly linear throughput scaling. BDLedger features a unique consensus mechanism called \textit{Random Witness Consensus} and uses a DAG-based block structure. To support this scaling, the system makes two key trade-offs: 1) it does not establish consensus on transaction order but focuses only on transaction content, which ensures that transactions do not conflict and can be processed in parallel; 2) it limits the nodes that store block content to a fixed number of random nodes, which helps control storage and bandwidth costs. The study demonstrates that BDLedger can scale throughput from approximately 20,000 TPS with 10 nodes to about 160,000 TPS with 100 nodes.

\cite{Fair_Assiri_2020} focuses on improving PBFT, a consensus-based algorithm used in the Tendermint blockchain~\cite{Kwon_tenderminttendermint_2024}. The paper introduces a lock-free algorithm for the blockchain, enhancing concurrency by allowing the proposal and voting phases to occur concurrently while keeping the commit phase sequential. This modification aims to mitigate the negative impacts of locks, thereby enabling greater parallelism within the blockchain system.

\cite{Parallel_Bai_2021} focuses on concurrency in the consensus algorithm level to avoid a throughput decrease by node numbers. They have proposed a more sufficient consensus algorithm based on PoV (read more in \ref{sssec:Consensus_Protocols}) named \textit{Parallel Proof of Vote (PPoV)}. In this protocol, concurrency is integrated into the consensus cycle, enabling numerous nodes to generate blocks simultaneously. Its research findings and experimental data indicate that PPoV outperforms traditional BFT consensus by a factor of 2-5, particularly when the number of nodes ranges between 4 and 100.

\subsection{Category 2: Sharding}\label{ssec:category-sharding}

Sharding, originally a database technique for splitting large datasets into smaller parts across multiple storage systems, improves efficiency by reducing server load. In blockchain, sharding was first introduced in~\citeyear{Luu_Secure_2016}~\cite{Luu_Secure_2016} as a method to enhance scalability and transaction throughput. The protocol \textit{ELASTICO}, proposed for permissionless blockchains, scales transaction rates almost linearly with available computational power. It efficiently uses network messaging and tolerates Byzantine adversaries up to one-fourth of the computational power. ELASTICO achieves secure partitioning of the mining network into smaller committees, each handling a distinct set of transactions or \textit{shards}. It was the first secure sharding protocol designed to operate in the presence of Byzantine adversaries.

Since then, sharding in blockchain has become one of the most advanced and specialized areas in the context of concurrent executions on blockchain networks. Sharding, in particular, has been extensively examined in several key surveys, and readers are encouraged to consult these sources for in-depth analysis. This survey aims to address critical gaps in the concurrency domain, which remains underexplored. For a more detailed overview of recent developments in the specific area of sharding, see \cite{Yu_Survey_2020}, \cite{survey_Liu_2023}, \cite{Sharding_Hashim_2022}, \cite{AlMutar_Scalability_2024}, \cite{Barat_SoK_2024}, \cite{Wang_SoK_2019}, \cite{Quan_Recent_2024}.

\subsection{Category 3: Intra-block Concurrent Processing}\label{ssec:category-Intra-block}
This approach acknowledges the sequential execution of transactions within a block by miners or validators in modern cryptocurrency systems like Ethereum and aims to address the limitations of serial execution of both inter and intra-Smart Contracts, which result in inefficiencies in system throughput and resource utilization. 

\subsubsection{\textbf{Speculative}}\label{sssec:category1:speculative}
To overcome this challenge, \cite{Adding_Dickerson_2020} presented a novel approach to enable the parallel execution of smart contracts by miners and validators. This approach utilizes established concepts from \citeauthor{Herlihy_Software_2003}'s distributed computing approaches \textit{software transactional memory (STM)}~\cite{Herlihy_Software_2003} and \textit{transactional boosting}~\cite{Herlihy_Transactional_2008}. The speculative approach has two strategies. In the first strategy, "execute then order," the leader speculatively executes smart contracts (SCs) in any arbitrary order, generates a corresponding schedule, and then sends this schedule to the validators, who replay the transactions and vote on the execution outcome. In the second strategy, "order then execute," the leader pre-determines the execution order of the SCs. Validators then speculatively execute the SCs in parallel, but they ensure that the final outcome reflects the predetermined order.

The first one is "order then execute," in which the leader chooses an order in which the SCs should appear to execute. The validators speculatively execute the SCs but ensure that the result seems to be executed in that order. The second strategy is "execute then order," in which the leader executes the SCs speculatively in an arbitrary order, generates a schedule, and sends that schedule to the validators who replay and vote on that execution. These strategies are discussed in more detail in the following. 

\textbf{Order then execute.} Miners execute smart contracts \textit{speculatively} in parallel, allowing non-conflicting contracts to proceed concurrently and generating a serializable concurrent \textit{schedule} for a block's transactions. Miners encode this schedule as a deterministic \textit{fork-join program}, facilitating parallel re-execution by validators. Despite its efficacy in increasing transaction throughput, this approach is based on the sequential nature of transaction execution within a block and does not introduce any uncertainty in transaction executions. Experimental results demonstrate significant speedups for both miners and validators, with a 1.33x speedup for miners and a 1.69x speedup for validators achieved with just three concurrent threads. Additionally, the paper presents a prototype implementation and discusses the potential for future advancements in multithreading support and programming language design to further improve throughput and concurrency control in smart contract execution. Furthermore, the paper highlights the applicability of the proposed mechanisms in permissioned blockchain systems and outlines future research directions, including support for multithreading in the Ethereum virtual machine and advancements in programming language support for smart contracts to maximize throughput while avoiding concurrency pitfalls~\cite{Adding_Dickerson_2020}.

This approach has been experimentally evaluated in \cite{Saraph_empirical_2019} to represent the potential benefits of speculative techniques for the execution of Ethereum smart contracts in parallel, using historical data to estimate their effectiveness. Transaction traces from sampled blocks on the Ethereum blockchain are replayed over time using a simple speculative execution engine, with miners attempting to execute all transactions in parallel and rolling back those causing data conflicts. Validators follow the same schedule as miners. The findings reveal that even simple speculative strategies yield notable speed-ups, with estimated speed-ups starting at approximately 8-fold in 2016 and declining to about 2-fold by the end of 2017, as transaction traffic increased. Moreover, the study identifies that a small set of contracts is responsible for a significant portion of data conflicts resulting from speculative concurrent execution, named \textit{storage hot-spots}. Several observations emerge from the study, including the importance of distinguishing between reads and writes to reduce conflict rates, limited additional benefits from more aggressive speculative strategies, and the potential for accurate static conflict analysis to yield modest benefits. Furthermore, increasing the number of cores in the simulated virtual machine improved speed-ups, but with little improvement beyond 64 cores. The study suggests directions for future research, including incentivizing contracts that produce fewer data conflicts and extending the virtual machine to support common commutative operations. In conclusion, while simple speculative strategies can produce significant speed-ups, these diminish as transaction rates and conflict rates increase. The study underscores the need to focus on reducing conflict rates to further increase parallelism in Ethereum-style smart contract execution, potentially through improved recognition of commuting operations at the semantic level and investigating the effects of intrinsic data types on contention.

\textbf{Execute then order.}
\cite{Dickerson_Conflict_2019} addresses the challenge of integrating black-box concurrent data structures into existing STM frameworks, focusing on optimistic transactional usage. It introduces conflict abstractions to define data-structure conflicts in terms of commutativity separately from object implementation, enabling efficient cooperation with generic STM runtimes. Additionally, shadow speculation allows individual transactions to make private speculative updates to highly concurrent black-box objects, facilitating opaque speculative updates that can be later dropped or atomically applied. These concepts are realized in a new transactional system called ScalaProust, built on ScalaSTM, supporting objects of arbitrary abstract type and integrating with underlying STM conflict-detection mechanisms. The paper's contributions include conflict abstractions, shadow speculation, ScalaProust system implementation, and experimental validation demonstrating competitive scalability with existing specialized approaches. However, the described mechanisms currently do not support mixtures between transactional objects and STM-managed read/write operations. Future directions include integrating pessimistic and optimistic treatment of black-box ADTs with standard STM memory operations, improving performance, and utilizing conflict abstractions for STM support of "pure writes" and automatic verification techniques for synthesis.

\subsubsection{\textbf{Deterministic}}
\cite{Efficient_Xia_2023} presents a two-stage approach for efficiently executing Blockchain transactions in parallel by adopting deterministic concurrency control. The paper addresses the inefficiencies of generating dependency graphs, which are often time-consuming and require re-executing transactions multiple times. Instead of building a large dependency graph, the authors partition transactions into \textit{batches} to avoid inter-batch conflicts, thus reducing the complexity of computing partial orders. The proposed \textit{deterministic concurrency control (DVC)} method enables high parallelism by solving an equivalent MinVWC problem (minimum vertex weighted coloring)~\cite{Wang_Reduction_2020}, which helps in finding an approximate optimal partial order without the need for re-execution of transactions. The authors demonstrate the effectiveness of their approach through experimental results, which show a significant reduction in the costs of computing the partial order and achieving a high degree of parallelism compared to existing solutions. This work offers a promising solution to improve the efficiency of concurrent Blockchain transaction execution, especially in the context of modern hardware utilization.

\subsubsection{\textbf{Partitioning}}\label{ssec:category1-partitioning}
This approach proposes a scalable smart contract execution scheme for blockchain-based systems, aiming to overcome the limitations of serial processing and improve system throughput. Using the decentralized nature of blockchain technology, the scheme enables the parallel execution of multiple smart contracts, thus enhancing the system's capability to handle a large volume of transactions. The study carried out in \cite{Gao_Scalable_2017} approach employs a divide-and-conquer strategy, using a fair contract partition algorithm based on \textit{integer linear programming (ILP)} to partition smart contracts into subsets. Each subset is then randomly assigned to a group of users, named sub-committee. Participants within each sub-committee exclusively handle a portion of smart contracts, enabling the simultaneous execution of various smart contracts. The contributions of this work include the development of a scalable execution scheme, theoretical analyses demonstrating its correctness and security, and simulations showcasing its practicality. The study evaluates the efficiency of the proposed scheme using data from existing smart contract systems, demonstrating its scalability and improved efficiency compared to sequential processing. 
However, a potential limitation lies in the efficiency of the ILP solver, which may pose challenges for larger sets of smart contracts. Future research directions include exploring more efficient and scalable partitioning techniques without relying solely on ILP solvers.
This paper focuses on PoW and is grounded in data collected from Ethereum. However, it is important to acknowledge that Ethereum transitioned from PoW to PoS during ``the merge'' in September 2022. Therefore, this study conducted in \citeyear{Gao_Scalable_2017} requires re-evaluation or revision to align with the changes resulting from Ethereum's transition to PoS.

\subsubsection{\textbf{Intra-block Performance Comparison}}\label{ssec:Intra-block-P-Performance-Comparison}

\begin{table*}
\caption{Speed Up Comparisons.}
\label{tab:Speed_Up_Comparisons}
\centering
\begin{threeparttable}
\fontsize{10pt}{12pt}\selectfont
\resizebox{\textwidth}{!}{
\begin{tabular}{|l|l|l|l|l|l|l|}
\hline
\textbf{Reference} & \textbf{Latest Year Test} & \textbf{Availability} & \textbf{Evaluation Execution Environment} & \textbf{Cores} & \multicolumn{2}{c|}{\textbf{Speed Up Rate}} \\ \cline{6-7}
 & & & & & \textbf{Proposer} & \textbf{Attestor} \\ \hline

Conthereum~\cite{Conthereum_Zareh_2025}, ~\citeyear{Conthereum_Zareh_2025} & 2025 & Public\textsuperscript{\href{https://github.com/Conthereum/conthereum}{1}} &
\begin{tabular}[c]{@{}l@{}}
    Intel Core i5-1135G7 (2.40GHz), eight logical \\
    processors, 16.0 GB RAM, Windows 11 Pro
\end{tabular}
& 3 & 2.93 & 2.48 \\ \cline{5-7} 
& & & & 4 & 3.73 & 2.77 \\ \cline{5-7} 

& & & & 5 & 4.31 & 3.23 \\ \cline{5-7} 
& & & & 32 (0\%conflict) & 25.92 & 25.92 \\ \cline{5-7} 
& & & & 32 (15\%conflict) & 9.08 & 5.58 \\  
\hline

\cite{Adding_Dickerson_2020}, \citeyear{Adding_Dickerson_2020} & N/A & Not available & 
\begin{tabular}[c]{@{}l@{}}
    4-core Intel Xeon W3550 (3.07 GHz),\\
    12 GB RAM, Ubuntu 16 
\end{tabular}
& 3 + 1 for GC & 1.33 & 1.69 \\ \hline

\cite{Saraph_empirical_2019}, \citeyear{Saraph_empirical_2019} & 2017 & Not available & N/A 
& 16 & 1.13 & N/A \\ \cline{5-7} 
& & & & 64 & 2.26 & N/A \\ \hline

Block-STM~\cite{blockstm_gelashvili_2023}, 2023
& 2023 & public\textsuperscript{\href{https://github.com/danielxiangzl/Block-STM}{2}}&
\begin{tabular}[c]{@{}l@{}}
AmazonWeb Services c5a.16xlarge instance,\\ 
(AMD EPYC CPU and 128GB memory) Ubuntu 18.04
\end{tabular}

& 32 (low-contention) & 20 & - \\ \cline{5-7} 
& & & & 32 (high-contention) & 9 & - \\ \hline

ScalaProust~\cite{Dickerson_Conflict_2019}, \citeyear{Dickerson_Conflict_2019} & - & Public\textsuperscript{\href{https://github.com/ScalaProust/ScalaProust}{3}} & 
\begin{tabular}[c]{@{}l@{}} 
    Amazon EC2 m4.10xlarge instance,\\
    which has 40 vCPUs and 160 GB of RAM
\end{tabular}

& - & - & - \\ \hline

\cite{Efficient_Xia_2023}, \citeyear{Efficient_Xia_2023} & N/A & Private &
\begin{tabular}[c]{@{}l@{}} 
    4-core processor of 2.2 MHz,\\
    4 GB main memory
\end{tabular}
& 4 & 3 & N/A \\ \cline{5-7} 
& & & & 20 to 24 & 5 & N/A \\ \hline
\end{tabular}
}
\begin{tablenotes}
\item[] \textsuperscript{1} \url{https://github.com/Conthereum}, 
\\
\textsuperscript{2} \url{https://github.com/danielxiangzl/Block-STM}, \\
\textsuperscript{3} \url{https://github.com/ScalaProust/ScalaProust}
\end{tablenotes}
\end{threeparttable}
\end{table*}

Table~\ref{tab:Speed_Up_Comparisons} presents a comparative analysis of the peak performance of various concurrency approaches at the intra-block level in blockchain, sorted in descending order based on the year of presentation. The \texttt{Reference} column specifies the cited work along with the year of publication. \texttt{Latest Year Test} denotes the most recent year in which blockchain transaction characteristics were evaluated within the respective study. \texttt{Evaluation Execution Environment} describes the hardware and software configurations of the experimental environment, including processor specifications, memory capacity, and operating system, as these factors can significantly impact performance results. 

The \texttt{Cores} column represents the number of processor cores explicitly allocated for transaction scheduling and execution. GC in this column refers to the cores that are allocated for garbage collection (GC) or other system processes. The \texttt{Speed Up Rate} column reports the speedup achieved in each study, further categorized into two distinct metrics: \texttt{Proposer} and \texttt{Attestor}, denoted as an ordered pair (p, a) for comparative analysis. Since attestors must maintain transaction order within a block, their speedup values are inherently lower than those of proposers for any given row in the table. If a value, including attestor performance, is not reported in a study, it is denoted as \texttt{N/A} (Not Available). For studies that present results across multiple core configurations, this table includes only representative values that allow for a comprehensive comparison.

The performance trends indicate that Conthereum~\cite{Conthereum_Zareh_2025} demonstrates the highest recorded speedup, achieving (2.92, 2.27) with 3 cores, surpassing the results of ~\cite{Adding_Dickerson_2020}, which reported (1.33, 1.69) for the same core count. Notably, Conthereum's proposer speedup of 2.92 with just 3 cores exceeds the 16-core and 64-core speedups of \cite{Saraph_empirical_2019}, which reported 1.13 and 2.26, respectively, without attestor evaluation. Additionally, \cite{Efficient_Xia_2023} did not examine attestor performance but tested proposers ranging from 4 to 24 cores, reporting a speedup of 3 for 4 cores and 5 for 24 cores. These values are outperformed by Conthereum, which achieves a proposer speedup of 3.73 for 4 cores and 5.81 for only 9 cores. This comparative analysis highlights the efficiency of Conthereum in achieving superior speedup with fewer computational resources.

\subsection{Category 4: Directed Acyclic Graph (DAG) approaches}\label{ssec:category-DAG}
This category of solutions redefines the structural foundation of blockchain systems by replacing the traditional linear sequential chain of blocks with a \textit{Directed Acyclic Graph (DAG)}~\cite{thulasiraman_graphs_2011}. The promise of DAG-based blockchain systems is to enable fast confirmation (complete transactions within million seconds) and high scalability (attach transactions in parallel) without significantly compromising security. 

In DAG-based systems, transactions are represented as nodes, with directed edges indicating dependencies or confirmations between them. This approach eliminates the sequential bottlenecks inherent in conventional blockchains, such as Bitcoin's PoW and Ethereum's PoS, enabling parallel transaction processing and significantly enhancing scalability and concurrency. DAG allows users to validate prior transactions to confirm their own, reducing energy consumption and transaction costs. By improving throughput and reducing latency, DAG-based architectures are particularly well-suited for high-frequency or microtransaction environments.

Despite these advantages, DAG-based systems face significant challenges, such as ensuring network consistency, preventing double-spending attacks in low-activity networks, and maintaining resilience against attacks that exploit transaction dependencies. These limitations, combined with DAG's early stage of development and the absence of proven security guarantees, explain why some well-established blockchains do not use it. In particular, security is often sacrificed for speed, and the blockchain trilemma cannot be fully resolved by DAG alone.

Several well-known DAG-based systems include IOTA, which uses the Tangle where each transaction confirms two prior transactions for enhanced scalability; Nano, which employs a block-lattice structure for fast, fee-less transactions by allowing each account to maintain its own blockchain; and Hashgraph, which offers a DAG-based consensus protocol focused on fairness, security, and efficiency.

In conclusion, while DAG-based systems have become a well-established area of research in blockchain concurrency, this survey focuses on the broader landscape of concurrency in blockchain systems. Extensive literature on DAGs is covered in several key surveys, and readers are directed to those sources for in-depth analysis. This survey aims to fill critical gaps in the concurrency domain, which remains underexplored. For specialized surveys on specific concurrency aspects, we provide references, ensuring a comprehensive overview while guiding readers to relevant detailed studies. For further exploration of DAG-specific advancements, please refer to~\cite{Wang_SOK_DAG_2023}, ~\cite{Huan_Chain_2022}, ~\cite{Wei_Survey_2022}, \cite{Fu_survey_2021}, and\cite{Lu_Survey_2024}.

\subsection{Category 5: Semi-Centralized Solutions} \label{ssec:category-Semi-Centralized}

Semi-centralized solutions in blockchain technology aim to balance the benefits of decentralization with the efficiency of centralized control. These approaches often seek to enhance scalability, transaction throughput, and governance by introducing certain centralized elements while maintaining the core principles of blockchain.

An illustrative example of a semi-centralized approach is the RCANE architecture \cite{RCANE_Van_2018}. RCANE proposes a network of parallel blockchains managed under a semi-centralized framework. This design allows for concurrent processing of transactions across multiple chains, effectively distributing the load and enhancing overall system performance. The semi-centralized governance model facilitates coordinated decision-making and efficient management, addressing scalability challenges inherent in fully decentralized systems.

\subsection{Category 6: Off-chain solutions} \label{ssec:category-offchain}
This research category discussed an intrinsic restriction of the blockchain in executing complex smart contracts. This limitation is made by considering gas to execute each instruction of a smart contract (\sec{}~\ref{ssec:Background_Smart_Contract}) and a limited threshold of gas supported for any block. For instance, sorting 256 integers using the selection sort algorithm demands 17 million gas, exceeding the current block limit of 8 million. The quick sort also reaches this gas limit after processing 2,000 elements. Consequently, more sophisticated applications, such as decentralized blockchain oracles, become infeasible to execute under these constraints~\cite{ACE_Wust_2020}. These smart contracts are name \textit{complex}~\cite{ACE_Wust_2020} or Computationally Intensive Contracts (CIC)~\cite{YODA_Sourav_2018}. There are some solutions to increase these per-block execution limits for complex smart contracts, and one of them is using \textit{off-chain} mechanism. An off-chain execution model, allows contract issuers to delegate contract execution to a selected group of service providers, thus operating independently from the blockchain's consensus layer. This section of the survey encompasses the concurrent solutions in this category of architecture.

ACE (Asynchronous and Concurrent Execution of Complex Smart Contracts)~\cite{ACE_Wust_2020} facilitates the execution of more complex smart contracts on permissionless blockchains using off-chain and the main benefit of ACE compared to earlier solutions is its ability to let one contract securely invoke another contract, even when they are executed by different sets of service providers. The system's key innovations include a flexible trust model and a robust concurrency control protocol, making it a significant advancement over previous solutions. The evaluation results show that ACE can facilitate the development of smart contracts that are significantly more complex than those on standard Ethereum.

In this protocol, concurrency is integrated into the consensus cycle, enabling numerous nodes to generate blocks simultaneously. PoV, in contrast, is an incentive mechanism designed to reward those who create \textit{value}. The value is defined for any smart contract based on the smart contract token transaction volume. POV allocates more tokens to the owners of smart contracts with higher transaction volumes. POV incentivizes value creators and manages the supply of coins through an adjustable incentive coefficient algorithm. 
To enhance POV performance on permissioned blockchains, the authors proposed and developed \textit{Hypernet}, a rapid off-chain transaction system that allows transactions to be processed independently of the blockchain. It consists mainly of the client and server components. The Hypernet client sends transactions through a centralized server that does not require trust. The centralized server's role in Hypernet is merely to forward tamper-resistant transaction messages. Hypernet introduces the concepts of multi-signature address and contract address to facilitate secure and fast off-chain transactions. This solution is evaluated by generating random parameters to trigger contracts and recording the actual time taken to issue rewards. Their results indicate that the system incurs a loss of approximately 2\% when POV is implemented, and Hypernet achieves transaction speeds four times faster than traditional permissioned blockchain systems.

\cite{Slimchain_Xu_2021} introduces SlimChain, a novel blockchain system designed to enhance transaction scalability through off-chain storage and parallel processing. Emphasizing a stateless approach, SlimChain retains only short commitments of ledger states on-chain, while off-chain nodes handle transaction executions and data storage. New schemes are proposed for off-chain smart contract execution, on-chain transaction validation, and state commitment. Additionally, optimizations are included to minimize network transmissions, and a new sharding technique is introduced to further boost system scalability. Extensive experiments validate SlimChain's performance, demonstrating a reduction in on-chain storage requirements by 97\% to 99\% and an improvement in peak throughput by 1.4 to 15.6 times compared to existing systems.

\subsection{Category 7: Cross-chain solutions}\label{ssec:Cross-chain}
Cross-chain solutions constitute a distinct category in smart contract concurrency due to their role in enabling atomic transactions and state synchronization across isolated blockchain networks. Unlike single-chain concurrency models, these approaches must address heterogeneous consensus protocols, varying finality times, and trust-minimized bridging mechanisms-all while preserving atomicity and preventing double-spending. Foundational work by \cite{atomic_herlihy_2018} established the theoretical framework for cross-chain swaps, while later surveys (\cite{survey_belchior_2021, Survey_Han_2023, Survey_Mao_2023, overview_Wei_2022,Survey_Peter_2021}) systematized the landscape.
Key concurrency challenges include:  \begin{enumerate*} [label=\roman*)]
\item  race conditions during cross-chain contract execution due to delayed finality,  
\item  verification of atomicity in multi-chain transactions, and  
\item  composability risks when smart contracts span multiple VMs. 
\end{enumerate*}

\subsection{Category 8: Transaction Ordering Dependency(TOD) and Front-Running Attacks}\label{ssec:category-TOD}
This category focuses on race conditions resulting from different miners' decisions regarding transaction orders within persisted blocks. As discussed in Section~\ref{ssec:Background_Smart_Contract}, each block of the blockchain contains verified transactions. \textit{Transaction Ordering Dependency (TOD)} refers to the uncertainty concerning the state of a contract from the user's perspective within the blockchain system at the time of contract invocation. When multiple transactions invoking the same contract occur almost simultaneously and are included in one block, the miner arbitrarily determines their order. This can lead to uncertainty for users regarding the state of the contract at the time of their transaction's execution, depending on the miner's decision. It is important to note that not all smart contracts are sensitive to transaction orders. Consequently, contracts affected by transaction ordering are termed TOD contracts. TOD contracts are susceptible to both benign and malicious invocations, the former potentially resulting in unexpected outcomes and the latter being exploited for unfair profit or theft~\cite{Luu_Making_2016}. Malicious exploitation of TOD-vulnerable smart contracts is termed \textit{front-running attacks}, which originate on the Chicago Board Options Exchange (CBoE) in ~\citeyear{Front_Markham_1988}~\cite{Front_Markham_1988}. Its definition by the Securities Exchange Commission (SEC) in 1977 is as follows: "The practice of effecting an option transaction based upon nonpublic information regarding an impending block transaction in the underlying stock, to obtain a profit when the options market adjusts to the price at which the block trades~\cite{Report_Securities_1979}". In a blockchain system, front-running occurs when malicious entities exploit the visibility of pending transactions in the mempool to execute transactions for their own benefit at the expense of the original transaction owner. In this attack adversaries can influence the outcome of this transaction race to prioritize their transaction order in the block through participation in mining, offering higher \texttt{gasPrice} to incentivize miners, or collusion with other miners.

As a benign scenario example, consider a smart contract that publishes puzzles and rewards participants for correct solutions. Suppose the contract owner can alter the reward amount by a contract invocation. In such cases, when a participant submits their solution transaction, there is no certainty they will receive the current prize amount if their solution is correct. This uncertainty arises because the owner may simultaneously submit a transaction to change the prize amount, and depending on the miner's decision, the participant may receive either the old or updated reward. Another example is a decentralized exchange or marketplace allowing users to buy and sell tokens, with fluctuating price functions callable by token owners. If a buy order and a price change transaction are submitted nearly simultaneously and included in the same block, the buyer cannot be certain whether they will purchase the token at the original or updated price, depending on the transaction order.

In malicious scenarios, the owner of a puzzle smart contract could monitor the network and, upon observing a solution being submitted, quickly send a transaction to change the reward amount to zero within the block interval (explained in~\ref{ssec:Background_Smart_Contract}), increasing the likelihood of their transaction being included before the solution transaction in the subsequent block. Consequently, the adversary can exploit the system to acquire puzzle solutions at minimal cost.

Detecting TOD involves recognizing valid transactions that influence a contract's global or state variables, akin to identifying read-after-write dependencies in race detection, which poses challenges for developers. The study conducted in~\cite{TODLER_Munir_2023} identifies different types of TODs, including a previously undocumented variant. To address TOD detection, the paper proposes TODler, a static analyzer based on information flow analysis. Evaluation of 108 Ethereum smart contracts demonstrates that TODler surpasses existing methods in terms of both speed and accuracy, while also identifying the newly discovered TOD pattern.

In \cite{TransRacer_Ma_2023} the authors introduce TransRacer, an innovative tool designed to detect transaction races in Ethereum smart contracts. Transaction races, which can result from the concurrent execution of transactions within the same block, have been largely overlooked despite their potential to cause inconsistent states and other unexpected outcomes. TransRacer employs symbolic execution to analyze function dependencies, enabling it to identify and generate witness transactions that reveal hidden races in specific contract states. This method significantly narrows the search space compared to random fuzzing techniques, enhancing detection accuracy and efficiency. Experimental results on 50 real-world smart contracts demonstrated TransRacer's capability to detect 426 races within approximately 256 minutes, including 149 significant race bugs. Further empirical analysis on 6,943 smart contracts underscores the prevalence and potential harm of transaction races in practice, highlighting the critical need for tools like TransRacer to maintain smart contract integrity.

\cite{Simulation_Stucke_2022} presents an open-source simulation tool designed to educate users on the vulnerabilities and mitigations associated with front-running attacks in Ethereum. This work introduces a comprehensive simulation environment that allows users to perform and understand various types of front-running attacks, such as displacement attacks, sandwich attacks, and priority gas auctions. The tool also demonstrates how to mitigate these attacks using the MEV-geth protocol, enhancing transaction privacy. By providing a hands-on educational platform, the simulation software aims to bridge the gap in educational resources regarding blockchain security, particularly in teaching smart contract vulnerabilities and mitigation strategies. This tool has been integrated into the curriculum at the University of Bristol, underscoring its utility in training the next generation of blockchain developers. The simulation's experimental framework highlights the practical challenges and potential solutions in managing front-running attacks, contributing valuable insights into the security dynamics of blockchain transactions.

\subsection{Category 9: Available concurrent transaction execution vulnerabilities (misconception)}\label{ssec:category-misconception}

In \citeyear{Sergey_Concurrent_2017}, \cite{Sergey_Concurrent_2017} introduced a concurrent perspective on smart contracts. It explained the similarities between the behaviors of smart contracts within cryptocurrency frameworks and the classic challenges encountered in shared-memory concurrency scenarios. By explaining two instances sourced from the Ethereum blockchain, the authors explained vulnerabilities to bugs that mirror those commonly encountered in traditional concurrent programs. In elaborating on these examples, the paper underscores the intrinsic relationship between observable contract behaviors and well-established concepts in concurrency, including atomicity, interference, synchronization, and resource ownership.

This study analogized smart contract utilization in blockchains to threads that engage with concurrent objects in shared memory (\textit{contracts-as-concurrent-objects}). Notably, the paper critiques the prevalent non-modular design of locking contracts, advocating instead for the implementation of synchronization primitives as standalone libraries. This proposed shift aligns with established software engineering best practices, where such modularization fosters better reasoning about contract behaviors. However, the separation of contract logic from locking mechanisms presents its own set of challenges, necessitating a holistic approach to contract verification that considers interactions with rigorously specified external contracts.

Moreover, the discourse extends to address concerns regarding liveness properties in contract implementations involving locks and exclusive access. By contemplating scenarios where certain accounts may indefinitely retain access, impeding progress and fairness within the system, the paper raises pertinent questions about ensuring eventual progress and incentivizing parties to release locks. This discussion underscores the importance of fairness assumptions akin to those found in concurrency theory and presents avenues for leveraging existing proof methods for reasoning about progress and termination in multi-contract executions.

Concluding with reflections on the broader implications of the proposed analogy, the paper advocates the use of insights from concurrency research to enhance understanding, debugging, and verifying complex contract behaviors within distributed ledger environments. Although acknowledging the limitations and unique aspects of contract programming, such as gas-bounded executions and fund management, the authors invite speculation on potential challenges and insights inspired by the observed parallels. These speculations range from exploring scenarios reminiscent of the ABA problem in non-garbage-collected languages to considering the influence of mining protocols on scheduling priorities and defining consistency notions for composite contracts with multi-transactional operations. Through this comprehensive examination, the paper underscores the interdisciplinary nature of blockchain research and its potential to benefit from insights gleaned from established fields such as concurrency theory.

\cite{Qu_Formal_2018} discusses the vulnerability of smart contracts, with a particular focus on concurrency related issues. The study highlights the critical challenges posed by the concurrent nature of smart contracts, which share similarities with concurrent programs with shared memory. Using formal verification methods, specifically the \textit{Communicating Sequence Processes (CSP)} theory~\cite{Hoare_CSP_1978} and the \textit{Failure Divergence Refinement (FDR)} model checking tool, the research aims to address these challenges. The paper illustrates a race condition in a smart contract example borrowed from Solidity language documents, named The Safe Remote Purchase smart contract~\cite{noauthor_solidity_2024}, alongside an attack scenario resulting from specific concurrent executions of this smart contract. The authors utilize CSP theory to formally model this smart contract and the considered attack, subsequently using the FDR model checker to demonstrate the possibility of this attack occurring. The findings underscore the potential advantages of using CSP and FDR tools for vulnerability detection within smart contracts, particularly in the context of concurrency.

However, our analysis reveals an error in the original assumption made in this paper regarding the race condition and the designed attack in smart contracts, specifically within Ethereum which is the subject of this study. The truth is that inherently, Ethereum's infrastructure does not support concurrency, contrary to the assumption made in this study. The concurrency assumption in \cite{Qu_Formal_2018} is based on the study in \cite{Sergey_Concurrent_2017}, which has led to ongoing debate despite its flaws. Consequently, it underscores the importance of rectifying such misconceptions within this branch of the body of literature in this field; the next section discusses this in more detail.

\section{Addressing the Flawed Assumption (Category 9)}\label{sec:FlawedAssumption}

This section delves into an erroneous assumption identified in the investigation, categorized as Category 9, as elaborated in Section~\ref{ssec:category-misconception}. The assumption is the presence of concurrent executions within smart contracts, implying the possibility of race conditions in the execution of multiple transactions inside one smart contract in the Ethereum infrastructure. However, this section demonstrates why this assumption is invalid. The first subsection presents evidence from the Ethereum references to refute the notion of concurrent execution. References from Ethereum's architecture clearly outline the sequential and non-concurrent execution of transactions within each block that consequently prevent any concurrent executions inside one smart contract. The subsequent subsection explores an alternative perspective on the potential for concurrent execution through the mechanism of sharding.

\subsection{Evidence from blockchain design}
\textbf{Ethereum.} \citeauthor{buterin_ethereum_white_2013} explicitly stated the sequential execution of transactions on Ethereum white paper as ``Note that the state is
not encoded in the block in any way; it is purely an abstraction to be remembered by the validating node and can only be computed (securely) for any block starting from the genesis state and sequentially applying every transaction in every block'' \cite{buterin_ethereum_white_2013}.

\subsection{Sharding and potential for transactions concurrency issue}

Sharding is a scalability technique that partitions the blockchain state and transaction processing across multiple shards (databases). This allows for parallel processing of transactions, potentially enabling true concurrency in smart contracts.

\subsubsection {Sharding architecture}
In a sharded architecture, transactions are routed to their designated shard based on a sharding key (e.g., user address). This enables concurrent processing of transactions within different shards. However, challenges remain:
\begin{itemize}
    \item \textbf{Cross-Shard Transactions:} Transactions that interact with data in multiple shards require careful coordination to maintain consistency.
    \item \textbf{State Synchronization:} Sharding introduces the need for efficient mechanisms to keep all shard states consistent with the overall blockchain state.
    \item \textbf{Security Considerations:} Careful design is necessary to ensure the security of the sharded system and prevent attacks that exploit shard boundaries.  
\end{itemize}
\subsubsection {Potential benefits of sharding for concurrency}

Despite the challenges, sharding offers potential benefits for achieving true concurrency in smart contracts:
\begin{itemize}
  \item \textbf{Increased Throughput:} Parallel processing can significantly increase the number of transactions processed per second.
   \item \textbf{Improved Scalability:} Sharding can accommodate a growing number of users and transactions.
  \item \textbf{Enhanced Flexibility:} Developers can design smart contracts that take advantage of shard-specific functionalities.
\end{itemize}

\begin{table*}[]
\caption{Comparison}\label{tab:comparison}
\resizebox{\textwidth}{!}{%
\begin{tabular}{lllllll}

Year & Ref & \textbf{Miner Approach} & \textbf{Locks} & \textbf{Require Block Graph} & \textbf{Validator Approach} & \textbf{Blockchain Type} \\

\citeyear{Adding_Dickerson_2020} & \citeauthor{Adding_Dickerson_2020}~\cite{Adding_Dickerson_2020} & Pessimistic ScalaSTM & Yes & Yes & Fork-join & Permissionless\\

\citeyear{Enabling_Zhang_2018} & \citeauthor{Enabling_Zhang_2018}~\cite{Enabling_Zhang_2018} & - & - & Read, Write Set & MVTO Approach & Permissionless\\

\citeyear{Anjana_EfficientFramework_2019} & \citeauthor{Anjana_EfficientFramework_2019}~\cite{Anjana_EfficientFramework_2019} & Optimistic RWSTM & No & Yes & Decentralized & Permissionless \\ 

\citeyear{Amiri_ParBlockchain_2019} & \citeauthor{Amiri_ParBlockchain_2019}~\cite{Amiri_ParBlockchain_2019} & Static Analysis& - & Yes & - & Permissioned\\

\citeyear{Saraph_empirical_2019} & \citeauthor{Saraph_empirical_2019}~\cite{Saraph_empirical_2019} & Bin-based Approach & Yes & No & Bin-based & Permissionless \\

\citeyear{Anjana_Entitling_2019} & \citeauthor{Anjana_Entitling_2019}~\cite{Anjana_Entitling_2019} & & & & & \\

\citeyear{Anjana_EfficientConcurrent_2021} & \citeauthor{Anjana_EfficientConcurrent_2021}~\cite{Anjana_EfficientConcurrent_2021} & Optimistic ObjectSTM & No & Yes & Decentralized & Permissionless \\

\citeyear{Anjana_EfficientParallel_2021} & \citeauthor{Anjana_EfficientParallel_2021}~\cite{Anjana_EfficientParallel_2021} & & & & & \\

\citeyear{DiPETrans_Baheti_2022} & \citeauthor{DiPETrans_Baheti_2022}~\cite{DiPETrans_Baheti_2022} & & & & & \\

\citeyear{DAGBased_Piduguralla_2023} & \citeauthor{DAGBased_Piduguralla_2023}~\cite{DAGBased_Piduguralla_2023} & & & & & \\

\citeyear{OptSmart_Anjana_2024} & \citeauthor{OptSmart_Anjana_2024}~\cite{OptSmart_Anjana_2024} & Bin+Optimistic RWSTM & No & No (if no dependencies)/Yes  & Decentralized & Permissionless \\
 
\end{tabular}%
}
\end{table*}

\section{Conclusion and future work}\label{sec:Conclusion_And_Future_Work}
In this paper, we have critically examined the assumptions surrounding concurrency in smart contracts, identifying misconceptions in the discourse on concurrent transaction execution. Through a comprehensive analysis of existing literature and concepts such as sharding, we have challenged prevailing paradigms and highlighted the need for a nuanced understanding of concurrency in blockchain systems. Moving forward, we advocate for continued research and development efforts to address the inherent trade-offs between scalability, security, and decentralization in blockchain networks.

\section*{Acknowledgments}
Work partially funded by the European Union under NextGenerationEU. PRIN 2022 Prot. n. 202297YF75, by the European Union under  NextGenerationEU, NRRP MUR program  FAIR - Future AI Research (PE00000013), and by the MUR PRIN 2020 - RIPER - Resilient AI-Based Self-Programming and Strategic Reasoning -\\ CUP E63C22000400001. Funded by the European Union. Views and opinions expressed are however those of the author(s) only and do not necessarily reflect those of the European Union or the European Commission. Neither the European Union nor the granting authority can be held responsible for them. Prof. Marco Roveri is partially funded by HE - CROSSCON - GA  101070537.

\bibliographystyle{plainnat}
\bibliography{references}
Abbreviations
The following abbreviations are used in this manuscript:

\begin{table}[]
\caption{Abbreviations}
\label{tab:Abbreviations}
\begin{tabular}{ll}
\hline
AUs  & Atomic Units                        \\
BFT  & Byzantine Fault-Tolerant            \\
BG   & Block Graph                         \\
CIC  & Computationally Intensive Contracts \\
CSP  & Communicating Sequence Processes    \\
DAG  & Directed Acyclic Graph              \\
DPoS & Delegated Proof of Stake            \\
DVC  & deterministic concurrency control   \\
ET   & Energy Trading                      \\
EVM  & Ethereum Virtual Machine            \\
FDR  & Failure Divergence Refinement       \\
ILP  & Integer Linear Programming          \\
Mempool  &  Memory Pool                    \\
MEV  & Maximal Extractable Value           \\
P2P  & Peer to Peer                        \\
PBFT & Practical Byzantine Fault Tolerance \\
PoA  & Proof of Authority                  \\
PoS  & Proof of Stake                      \\
PoV  & Proof of Value                      \\
PoV  & Proof of Vote                       \\
PoW  & Proof of Work                       \\
PPoV  & Parallel Proof of Vote             \\
SC   & Smart Contract                      \\
SMuLC & Sequential Multi-Layer Consensus   \\
STM  & Software Transactional Memory       \\
TOD  & Transaction Ordering Dependency     \\
TPS  & Transaction Per Second              \\
TTP  & Trusted Third Party                 \\ \hline
\end{tabular}
\end{table}
\section{About the Authors}

\noindent\begin{minipage}{\textwidth}
\begin{wrapfigure}[9]{l}{1in}
\vspace{-0.75ex} 
\includegraphics[width=1in,height=1.25in,clip,keepaspectratio]{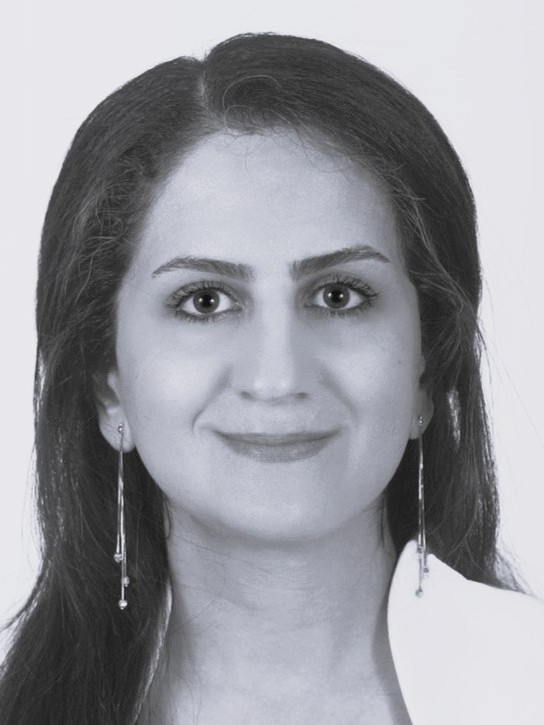}
\end{wrapfigure}
\noindent\textbf{Atefeh Zareh Chahoki} is a Ph.D.\ candidate in Computer Science at the University of Trento. 
Her research interests include blockchain system reliability, formal verification of smart contracts, and transaction scheduling optimization in distributed ledgers. 
She publishes in venues such as \textit{IEEE Transactions on Information Forensics and Security} and is committed to open science-releasing all implementations as open-source software. Before commencing her doctoral studies, she acquired extensive industry experience in enterprise software development, project management, and university lecturing.
\end{minipage}

\vspace{0.5\baselineskip}

\noindent\begin{minipage}{\textwidth}
\begin{wrapfigure}[9]{l}{1in}
\vspace{-0.75ex} 
\includegraphics[width=1in,height=1.25in,clip,keepaspectratio]{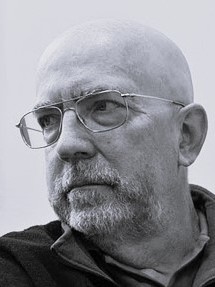}
\end{wrapfigure}
\noindent\textbf{Maurice Herlihy} has an A.B. in Mathematics from Harvard University, and a Ph.D. in Computer Science from M.I.T. He has served on the faculty of Carnegie Mellon University and the staff of DEC Cambridge Research Lab. He is the recipient of the 2003 Dijkstra Prize in Distributed Computing, the 2004 G$\ddot{o}$del Prize in theoretical computer science, the 2008 ISCA influential paper award, the 2012 Edsger W. Dijkstra Prize, and the 2013 Wallace McDowell award. He received a 2012 Fulbright Distinguished Chair in the Natural Sciences and Engineering Lecturing Fellowship, and he is fellow of the ACM, the National Academy of Inventors, the National Academy of Engineering, and the National Academy of Arts and Sciences. In 2022, he won his third Dijkstra Prize.
\end{minipage}

\vspace{0.5\baselineskip}

\noindent\begin{minipage}{\textwidth}
\begin{wrapfigure}[9]{l}{1in}
\vspace{-0.75ex} 
\includegraphics[width=1in,height=1.25in,clip,keepaspectratio]{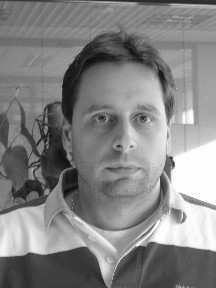}
\end{wrapfigure}
\noindent\textbf{Marco Roveri} is an Associate Professor in the Department of Information Engineering and Computer Science at the University of Trento, Italy. He was a Senior Researcher in the Embedded Systems Unit of Fondazione Bruno Kessler, and before a researcher in the Automated Reasoning Division of the Istituto Trentino di Cultura. His research interests include formal verification of hardware and software systems, formal cyber security, formal requirements validation, and automated model-based planning, and application of such techniques in industrial settings.
\end{minipage}

\end{document}